\documentclass[10pt,conference]{IEEEtran}

\usepackage{subcaption}
\usepackage{cite}
\usepackage{balance}
\usepackage{makecell}
\usepackage[T1]{fontenc}
\usepackage[utf8]{inputenc}
\usepackage{etoolbox}
\usepackage{url}
\usepackage{color}
\usepackage{verbatim}
\usepackage{siunitx}
\usepackage{multicol}
\usepackage{tabularx}
\usepackage{booktabs}
\usepackage{tablefootnote}
\usepackage[noend]{algpseudocode}
\usepackage{color,soul}
\usepackage{graphicx}
\usepackage{float}

\definecolor{_eddie}{RGB}{165,42,42}
\definecolor{_ruben}{RGB}{127,0,255}
\definecolor{_todo}{RGB}{153,0,76}
\definecolor{_spencer}{RGB}{255,128,0}
\definecolor{_mike}{RGB}{40,180,180}
\definecolor{_ali}{RGB}{203,38,38}
\definecolor{_josh}{RGB}{50,40,186}
\definecolor{_rock}{RGB}{153,76,0}
\definecolor{_notes}{RGB}{37,162,32}
\definecolor{_richelle}{RGB}{255,0,127}
\definecolor{_nirmalk}{RGB}{255,0,255}

\makeatletter
\def\BState{\State\hskip-\ALG@thistlm}
\makeatother
\begin{document}

\title{FPGA Based Emulation Environment for Neuromorphic Architectures}

\author{\IEEEauthorblockN{Spencer Valancius\IEEEauthorrefmark{1},
		Edward Richter\IEEEauthorrefmark{1},
		Ruben Purdy\IEEEauthorrefmark{1},
		Kris Rockowitz\IEEEauthorrefmark{1},
		Michael Inouye\IEEEauthorrefmark{1},
		Joshua Mack\IEEEauthorrefmark{1}, \\
		Nirmal Kumbhare\IEEEauthorrefmark{1},
		Kaitlin Fair\IEEEauthorrefmark{2}, 
		John Mixter\IEEEauthorrefmark{3} and 
		Ali Akoglu\IEEEauthorrefmark{1}} 
	\IEEEauthorblockA{\IEEEauthorrefmark{1}Department of Electrical and Computer Engineering, University of Arizona, Tucson, AZ, 85719 USA \\
		\{svalancius12, edwardrichter, rubenpurdy, rockowitzks, mikesinouye, jmack2545, nirmalk, akoglu\}@email.arizona.edu
	}
	\IEEEauthorblockA{\IEEEauthorrefmark{2}Air Force Research Labs, Florida, 32542 USA\\
		kaitlin.fair@us.af.mil}
	\IEEEauthorblockA{\IEEEauthorrefmark{3}Raytheon Missile Systems, Tucson, AZ, 85747 USA\\
		John\_E\_Mixter@raytheon.com}
}

\maketitle

\begin{abstract}
Neuromorphic architectures such as IBM’s TrueNorth and Intel’s Loihi have been introduced as platforms for energy efficient spiking neural network execution. However, there is no framework that allows for rapidly experimenting with neuromorphic architectures and studying the trade space on hardware performance and network accuracy. Fundamentally, this creates a barrier to entry for hardware designers looking to explore neuromorphic architectures.   In this paper we present an open-source FPGA based emulation environment for neuromorphic computing research. We prototype IBM’s TrueNorth architecture as a reference design and discuss FPGA specific design decisions made when implementing and integrating it’s core components. We conduct resource utilization analysis and realize a streaming-enabled TrueNorth architecture on the Zynq UltraScale+ MPSoC. 
We then perform functional verification by implementing networks for MNIST dataset and vector matrix multiplication (VMM) in our emulation environment and present an accuracy-based comparison based on the same networks generated using IBM's Compass simulation environment. 
We demonstrate the utility of our emulation environment for hardware designers and application engineers by altering the neuron behavior for VMM mapping, which is, to the best of our knowledge, not feasible with any other tool including IBM's Compass environment. The proposed parameterized and configurable emulation platform serves as a basis for expanding its features to support emerging architectures, studying hypothetical neuromorphic architectures, or rapidly converging to hardware configuration through incremental changes based on bottlenecks as they become apparent during application mapping process.


\end{abstract}

\section{Introduction}\label{sec: 01_INTRO}
 
Spiking neural network (SNN) architectures have been proposed with the goal of creating non-von Neumann architectures that emphasize the strengths of biologically inspired neural networks: low-power, high parallelism, and fast complex computations \cite{arXiv_2017_OtherPlatforms,IJCNN_2013_Foundation}. 
IBM's TrueNorth Chip \cite{TCAD_2015_Foundation} and Intel's Loihi~\cite{davies_loihi:_2018} are examples of such architectures for modeling leaky-integrate-and-fire neurons with the capability of implementing multiple types of dynamic and stochastic neuron models. 
Neuromorphic computing architectures have a number of configuration parameters that are not inherent to the hardware such as the number of weights a neuron can have, the bitwidth of these weights, synaptic weight memory and precision, synaptic delay, the number of neurons and axons in a core, number of cores, neuron count per core, network topology, and the constraints used during training networks for deployment onto the target neuromorphic architecture. 
There is a need for an open-source configurable emulation environment for hardware architects and application engineers to investigate performance bottlenecks and accordingly alter the architecture by investigating the impact of their design decisions on hardware performance through trend based analysis. 
Such design space exploration and prototyping based efforts are not feasible without an emulation environment as these neuromorphic chips are designed as ASICs.

In this study we present a parameterized and configurable emulation platform that serves as a basis for supporting other neuromorphic architectures or investigating new architectures targeted for different application domains. 
We recreate and implement the TrueNorth architecture as a reference design on the Xilinx Zynq UltraScale+ MPSoC ZCU102. 
We validate the functionality of our emulation environment using both the MNIST dataset and vector matrix multiplication as case studies. 
For the MNIST dataset we implement a network by Esser et al. \cite{NIPS_2015_Training} in our environment and compare the results of the two networks. 
For the case of vector matrix multiplication (VMM), we replicate VMM mapping method of Fair et al.~\cite{fair19} and compare the results against similar networks generated using IBM's Compass environment~\cite{Compass}. 
We then demonstrate the architectural prototyping capabilities of our environment by introducing a single change to the neuron block component. 
This alteration, without accuracy degradation to either the MNIST or VMM case studies, reduces the resource requirements of the VMM networks by 50\%.

Even though TrueNorth and Loihi architectures are different in terms of packet processing, SNN mapping, core architecture, and core synchronization, our emulation environment allows tuning aforementioned key configuration parameters to execute SNNs targeted for Loihi. 
The parameterized design enables the manipulation of core components, without the need to recreate the entire design from scratch. 
This allows for a user to selectively utilize TrueNorth components that interface with their own unique implementations, such as using the TrueNorth's crossbar but changing out the routing network for some other interconnect architecture. 
The modularity of our emulation environment allows for incremental changes to implement features such as on-chip learning, core-to-core multicast, core management packets, and configurable synaptic weight precision (signed and unsigned) offered by the Loihi.

The remainder of this paper is organized as follows: In Section \ref{sec: 02_TN_OVERVIEW} we introduce each individual component that create a single TrueNorth Core and describe our key FPGA implementation methods. 
In Section \ref{sec: 04_VERIFICATION}, we present hardware resource requirements of the proposed emulation environment, analyze its scalability on the FPGA and discuss our validation approach.
Finally, in Section \ref{sec: 05_CONCLUSION} we present our conclusions and planned future work.

\section{Reference Architecture Overview and Implementation}\label{sec: 02_TN_OVERVIEW}
\begin{figure}[]
    \centering
    \includegraphics[width=0.5\linewidth]{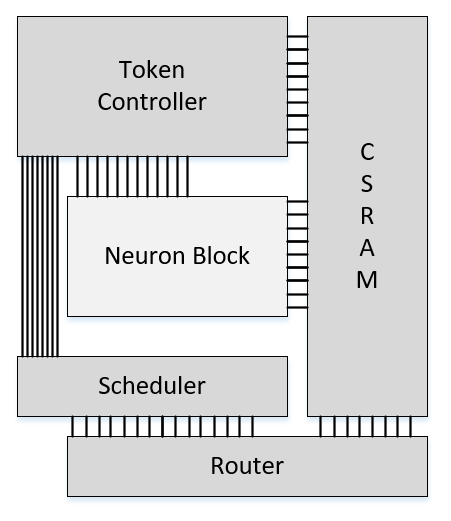}
    \caption{TrueNorth core comprised of  five components: \textit{neuron block}, \textit{core sram}, \textit{router}, \textit{scheduler}, and \textit{token controller}.}
    \label{fig: truenorth_archtecture}
    \vspace{-4mm}
\end{figure}

A single TrueNorth chip is comprised of 4,096 neurosynaptic cores.
Each core is comprised of five components~\cite{TCAD_2015_Foundation} as shown in Figure \ref{fig: truenorth_archtecture}.
In this section we summarize functionality of each component and describe our FPGA based implementation approach.  

\subsection{Neuron Block}\label{subsec: 2a_1_neuron_block}
The \textit{neuron block} is the primary computational component for our reference architecture.
The purpose of the \textit{neuron block} is to compute a running sum value known as the \textit{neuron potential}, which is the sum of the weights associated with the input axons, and accumulate it with the existing potential value.
Each of the individual neurons within a core are sequentially evaluated.
During a neuron's evaluation, all input axons are checked for a binary spike as well as a synaptic connection to the current neuron.
If both the binary spike on the axon, and a synaptic connection between the neuron and the axon exists, then a corresponding synaptic weight is added to the neuron's potential.
Once all input axons have been checked, a leak value is then applied to the potential.
This potential is then evaluated against an asymmetrical threshold value.
If the potential is greater-than or equal-to a positive threshold, then the \textit{neuron block} produces a spike, and the neuron potential is reset to a reset-potential value.
If the potential is less-than a negative threshold, then the \textit{neuron block} resets the neuron potential to a reset-potential value, but does not produce a spike.
In the event that neither of these cases occur, the neuron's potential is saved and used as the start of new running sum during the next cycle \cite{IJCNN_2012_Foundation_Core,IJCNN_2013_Foundation,IJCNN_2013_Foundation_Neuron,TCAD_2015_Foundation}.

\begin{figure}[t]
    \centering
    \includegraphics[width=1.0\linewidth]{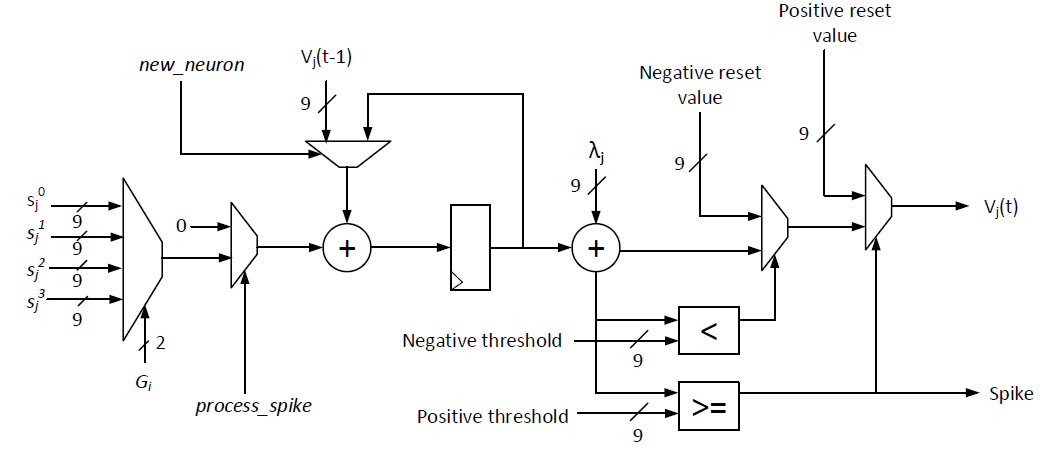}
    \caption{Neuron Block}
    \label{fig: neuron_block}
    \vspace{-4mm}
\end{figure}
For our emulation we produce a \textit{neuron block} model as shown in Figure \ref{fig: neuron_block}.
When a new neuron is loaded into the \textit{neuron block}, the four synaptic weights are loaded into the multiplexer unit on the far left of the image.
The synaptic weight is selected based on the current input axon's synaptic weight index value. 
This index value is held constant for a given axon.
For example, if input axon 0 has a synaptic weight index value of 3, that index value will be used to select the same index of synaptic weight from the mux for all neurons.
If, for a given input axon, there exists a binary spike as well as a synaptic connection, then the next mux will allow the selected synaptic weight through, and it will be added to the current neuron potential.
If either of these conditions is not met, then this mux will send the value of 0 through and no change will be made to the neuron's potential.

The updated neuron potential is saved in the register to be used on the next clock cycle for the next axon.
A leak value $\lambda$ is constantly incorporated into the updated neuron potential to allow for the modification of the potential value outside of spikes.
The leak modified neuron potential is then compared against the positive and negative thresholds to determine if a spike is to be processed as well as which reset-value, if any, to utilize.
Once the current neuron has been evaluated, a new neuron's neuron potential is used for the running sum, rather than continuing with the old neuron's value.

\subsection{Core SRAM}\label{subsec: 2a_2_core_sram}
The \textit{core sram} is the main memory for a TrueNorth core.
Each \textit{core sram} is a matrix of 256 rows by $X$ columns where $X$ is the total number of bits required to fully encode a single neuron with all of its respective parameters.
Our \textit{core sram} model is very much like the one described by Akopyan et al.~\cite{TCAD_2015_Foundation}. In our implementation $X$ is equal to $386$ bits and is broken down as shown in Table \ref{tab: csram_breakdown}.
In a \textit{core sram}, each of the 256 rows represents one of the 256 neurons contained within a single TrueNorth core.

\begin{table}[t]
\caption{Core SRAM Parameter Breakdown.}
\centering
\scalebox{1}{\begin{tabular}{l l}
\toprule
Parameter Name & Bit Width \\
\midrule
Synaptic Connections & 256 bits \\
Potential and Neuron Parameters & 100 bits \\
Spike Destination & 26 bits \\
Delivery Tick & 4 bits \\
\midrule
Potential \& Neuron Parameters & Bit Width \\
\midrule
Current Potential & 9 bits \\
Reset Potential & 9 bits \\
Weights 0 1 2 and 3 & 9 bits each \\
Leak Value & 9 bits \\
Positive/Negative Threshold & 18 bits each \\
Reset Mode & 1 bit \\
\bottomrule
\end{tabular}}
\vspace{-12pt}
\label{tab: csram_breakdown}
\end{table}

For our emulation's \textit{core sram} component, we implement it as two different modules: controller and memory.
The controller determines which row, or rather which neuron, in the memory module to read from, as well as notifies the token controller when all rows have been processed.
By separating the controller from the memory module, we assist the synthesis tool in mapping the memory module to the FPGA's BRAM more efficiently.
This results with implementing the \textit{core sram} module with only 5.5 BRAM blocks per core, which would have required 386 BRAM blocks otherwise.

Each neuron accounts for 386 bits of information, therefore each emulated core requires 98,816 bits ($256*386$) to store all the neuron information. 
However, 5.5 BRAM elements offers 202,752 bits of memory, which is much more than the 98,816 bits that we are storing for a single \textit{core sram}. 
Despite using only around 50\% of the total memory storage available in the provided 5.5 BRAM blocks, our \textit{core sram} needs to be able to read all 386 bits for a single neuron in a single clock cycle.
Our \textit{core sram} uses 5 36-Kb BRAM blocks in a $512\times72$ configuration and a single 18-Kb BRAM block in a $512\times36$ configuration to match this need.

\subsection{Router}\label{subsec: 2a_3_router}

\begin{figure}[t]
\begin{subfigure}{.5\textwidth}
\centering
\includegraphics[width=0.9\linewidth]{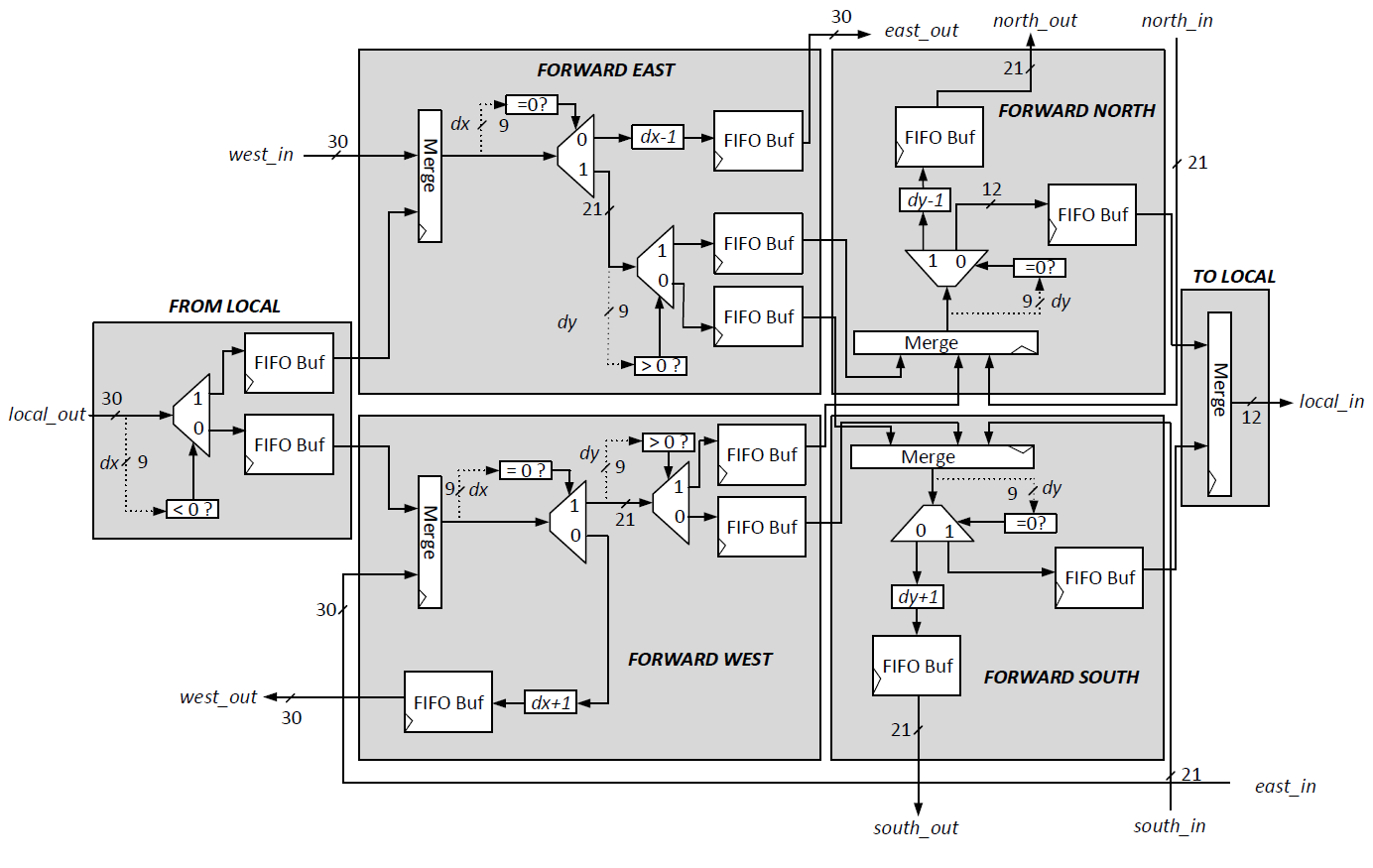}
\caption{Our Router}
\label{fig:ourrouter}
\end{subfigure}

\begin{subfigure}{.5\textwidth}
\centering
\includegraphics[width=0.9\linewidth]{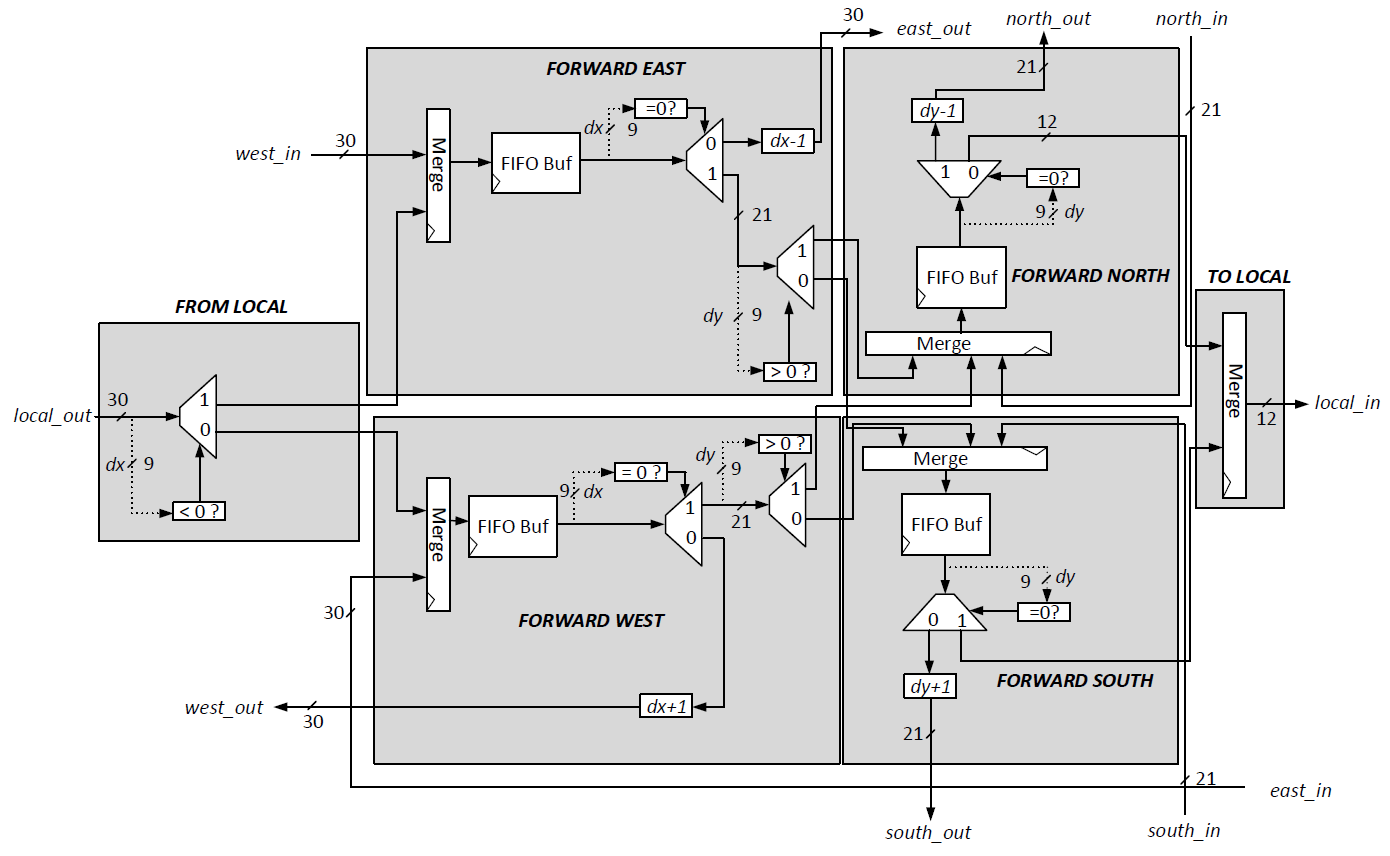}
\caption{IBM TrueNorth Router}
\label{fig:ibmrouter}
\end{subfigure}
\caption{Differences between our router and TrueNorth router:   Synchronous design and reorganizing buffers by moving them after the merge simplified   back-pressure logic and increased our ability to have high throughput in times of high congestion.}
\label{fig:bothrouters}
\vspace{-4mm}
\end{figure}

The \textit{router} is responsible for the inter-core communication in the TrueNorth chip and enables delivering spike packets between adjacent cores from source neurons to destination axons.
Each spike in the TrueNorth network is represented as a packet, which contains information regarding number of cores to travel horizontally ($\Delta x$) [9 bits] and vertically ($\Delta y$) [9 bits], which axon in the core to be delivered to [8 bits], and which tick to be delivered on [4 bits]. Packets travel first horizontally, and then vertically 
across the two-dimensional mesh of cores until they arrive at the destination core, where they are sent to the scheduler. 
At the final destination, the scheduler uses the remaining bit values to determine which axon and \textit{tick} instance to save the spike to.

Figure \ref{fig:ourrouter} shows our implementation of the \textit{router} component. 
Each router has forward east, forward west, forward north, forward south, from local, and to local modules, which are used to communicate with both the internal modules of the core as well as the adjacent cores in every cardinal direction. 
Forward east, forward west, forward north, and forward south each have one to three FIFO buffers which are necessary for two reasons. 
First, when mapping a non-trivial application to TrueNorth, the number of packets simultaneously traveling through the network on-chip can be significantly large. 
At times of high congestion, the buffers are necessary to achieve high throughput. 
Second, as packets travel through the  routing network, the FIFO buffers that capture packets within each forward module become full. 
This is addressed by applying a form of back-pressure into the network as discussed by Akopyan et al. \cite{TCAD_2015_Foundation}. 
Buffers enable back-pressure to ensure that packets are not lost when traveling through the network.

A significant difference between our implementation and the original router implementation in IBM's TrueNorth shown in Figure~\ref{fig:ibmrouter} is the placement and number of buffers.
In the original implementation, each cardinal direction has a single buffer to store inputs arriving at that forwarding direction.
This setup requires complicated backpressure logic as each buffer can send packets in multiple directions. 
For example, the buffers in forward east can go to forward north, forward south, or to the eastern core's forward east module. 
Therefore, in order to implement backpressure, the buffer needs to receive feedback from these three possible locations. 
Our implementation increases the number of buffers.
Rather than having a single buffer that buffers the input into the forward module, we have a buffer for each module's output.
This places each buffer in-between two merges (except for the two buffers in from local). 
The merge at the output of the buffer will send a \textit{read\_enable} signal to the buffer when the buffer is not empty and the buffers at the output of the merge are not full. 
Each buffer sends a \textit{buffer\_full} signal to the merge at its input to ensure that new data will not arrive when the buffer is full. 
This reduces the logic required to implement the router's backpressure. 
Additionally, by buffering the outputs of a forward module over buffering the inputs, we obtain a throughput increase of 3x for horizontal forwarding modules and 2x for vertical forwarding modules.

\subsection{Scheduler}\label{subsec: 2a_4_scheduler}
\begin{figure}[t]
    \centering
    \includegraphics[width=1.0\linewidth]{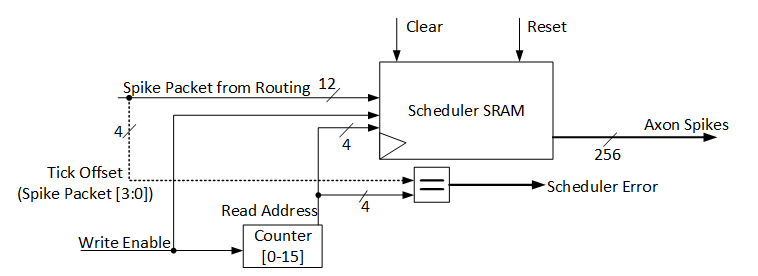}
    \caption{A synchronous scheduler design replaces 16 asynchronous control blocks with a counter.}
    \label{fig:scheduler}
\vspace{-4mm}
\end{figure}

The \textit{scheduler} is the final stop for a spike packet that is traversing the cores through the routing network.
By the time a spike packet arrives at its destination core, it is reduced to 12 bits  with 8 bits of axon and 4 bits of tick offset values. 
The \textit{scheduler} contains a 256 row by 16 column SRAM.
The 8 bit axon value of the spike packet is used to determine which of the 256 rows the newly arrived spike will be written on.
The 4 bit tick offset is used to determine which column, with respect to the currently active column, the spike is written to.
In the event that the tick offset causes the spike to be written to the active column, or the current \textit{tick}, an error is thrown and the packet is dropped.
This error does not cause any part of the TrueNorth core to halt.
It is %
used to alert the user that information may no longer be accurate \cite{TCAD_2015_Foundation}.

The scheduler in the reference design is comprised of the SRAM memory and control blocks. 
The SRAM memory is used to store the spike packets that arrive from the core's router. 
The control block determines which spikes to process in a given \textit{tick} instance. 
The \textit{scheduler} contains sixteen of these control blocks corresponding to each of the 16 columns (16 \textit{tick} instances), in which only one of these control blocks is active at a time by means of a passing token. 
The active control block reads the corresponding SRAM column and sends the data and the input axon spike information to the token controller. 
It is also responsible for then clearing the column once the token controller has completed its full FSM circuit and is back to waiting for the next \textit{tick}.

In our emulation design, we propose replacing the control logic with a 4-bit counter that increments every time it receives the tick as illustrated in Figure~\ref{fig:scheduler}. 
As the counter updates based on a signal received from the token controller, the counter value is sent into the look up table (LUT) memory blocks that comprise the \textit{scheduler} SRAM. 
We purposefully use LUT memory for the \textit{scheduler} over using additional BRAM blocks so that we may use the BRAM blocks wholly for the core SRAM components. 
As with the original scheduler design in TrueNorth, our emulation \textit{scheduler} does not allow spike packets to be written to a SRAM column corresponding to the current \textit{tick} instance. 
In TrueNorth's scheduler, if this occurs the packet is dropped and an error flag is sent to the token controller, which causes another flag to be raised to alert the user that information may no longer be valid. 
For our own emulation design we take this error flag and bypass the token controller, sending it out to the user. 
This allows us to distinguish between an error that has occurred within the scheduler, and an error within the token controller. 
Being able to distinguish between these two errors is a key item for our emulation environment, as new architecture prototyping may cause a distinct error in only one of these locations. 
Knowing which location can assist researchers in the debugging process.

\begin{table*}[t]
\caption{Post implementation resource usage by component for 1 and 5 core networks on Xilinx Zynq Ultrascale+ XCZU9EG FPGA based on Look Up Table Logic (LUTs), Look Up Table Memory(LUT-RAM), Block Random Access Memory (BRAM).}
\centering
    \begin{tabular}{|r|c|c|c|c|c|c|c|c|c|}
    \hline
    \multicolumn{1}{|c|}{Component}       & \multicolumn{2}{c|}{LUTs}    & \multicolumn{2}{c|}{LUT-RAM}    & \multicolumn{2}{c|}{FFs}   & \multicolumn{2}{c|}{BRAM} & Critical Delay        \\ \hline
    Network Size                          & 1            & 5            & 1            & 5            & 1            & 5            & 1           & 5           &   1                    \\ \hline
    Core SRAM                                 & 91          & 455          & 0            & 0            & 0            & 0           & 5.5         & 27.5        & 3.338 ns              \\
    Neuron Block                                    & 39           & 195         & 0            & 0            & 9            & 45            & 0           & 0           & 1.670 ns              \\
    Token Controller                                    & 46           & 230          & 0            & 0            & 32           & 160          & 0           & 0           & 5.321 ns              \\
    Scheduler                             & 376           & 1880          & 304          & 1520         & 4            & 20           & 0           & 0           & 4.647 ns              \\
    Router                                & 1418          & 7862         & 0            & 0            & 1167          & 2696         & 0           & 0           & 11.551 ns                \\
    \hline
    \multicolumn{1}{|l|}{Total Available} & \multicolumn{2}{c|}{274080} & \multicolumn{2}{c|}{144000} & \multicolumn{2}{c|}{548160} & \multicolumn{2}{c|}{912} & \multicolumn{1}{l|}{} \\ \hline
    \end{tabular}
\label{tab: Resource_Requirements_1and5}
\vspace{-8pt}
\end{table*}

\subsection{Token Controller}\label{subsec: 2a_5_token_controller}
\begin{figure}[t]
    \centering
    \includegraphics[width=1.0\linewidth]{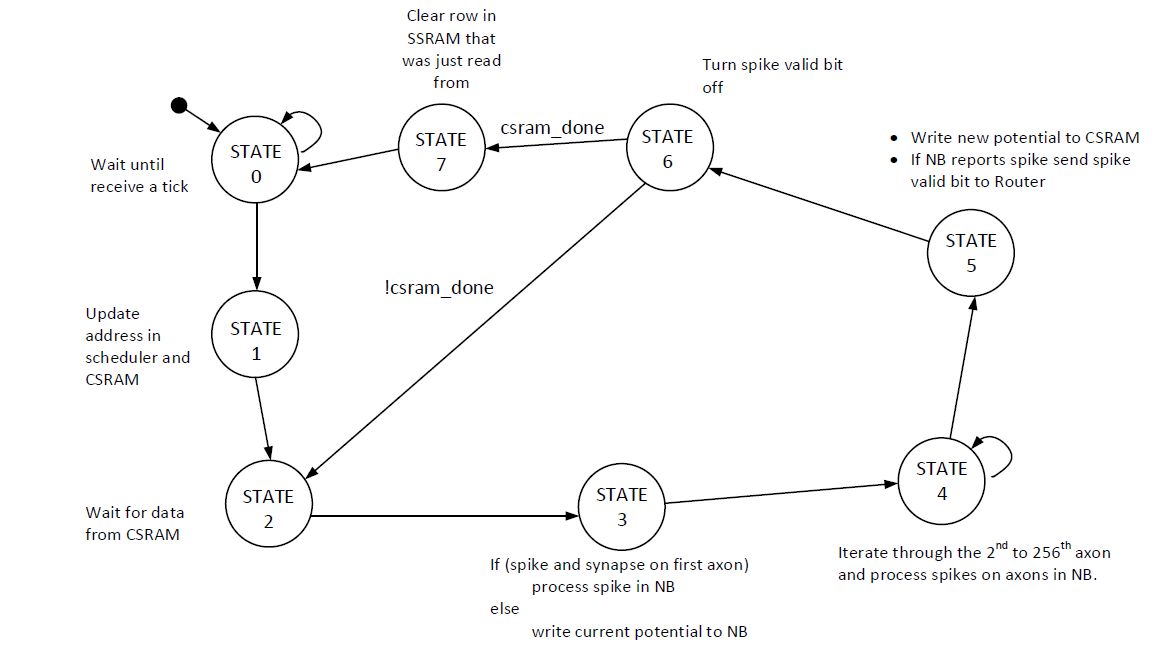}
    \caption{Token controller state machine.}
    \label{fig:tokencontroller}
    \vspace{-4mm}
\end{figure}

The \textit{token controller} maintains global synchronous behavior of the TrueNorth core, as well as intercommunication between the other four components in a single core.
The token controller is a 269 state asynchronous FSM in the original design.
256 states are used to evaluate the individual input axons to determine if a binary spike as well as a connection to the current neuron exists.
In the event this is true, the same state sends the appropriate information to the neuron block and uses an asynchronous communication tree referred to as the \textit{request/acknowledge tree} to meet timing constraints.
If an axon does not have either a binary spike or a synaptic connection with the current neuron then it moves on to the next state in the FSM.
The remaining 13 states reference the active neuron's information from the \textit{core sram} for the purpose of setting the current \textit{tick} instance within the scheduler and to retrieve the input axon binary spikes from the scheduler sram. A spike packet is sent to the router if the core determines the neuron block has met its defined threshold. The controller FSM then checks if all neurons have been evaluated.

In our design we take advantage of our fully synchronous design to reduce the number of states from 269 to 8 as shown in Figure~\ref{fig:tokencontroller}. 
We are able to do this by collapsing the 256 states used to evaluate the input axons into only a single state that loops for the number of input axons in our core. 
We are able to achieve this state reduction by first collapsing the 256 states used in TrueNorth for the input axon evaluation into a single state loop (state 4 in Figure~\ref{fig:tokencontroller}), where the number of loop iterations is equal to the number of input axons of 256.
Among the remaining thirteen states used in the TrueNorth token controller, we merge five states that deal with address updating to the scheduler and core sram components (state 1), and merge two states that deal with updating the current neuron's potential into the core sram and sending the spike information to the router (state 5). 
We re-purpose the spike debugging state in the original TrueNorth design to instead switch off the valid bit signal sent to our router after we deliver a spike packet from the current neuron (state 6).
For our first and last state we retain their original uses as described by TrueNorth. Lastly, we are able to remove three states from the TrueNorth design, as they correspond to states that don't impact the overall behavior of our emulation design.

\subsection{Output Buffer}\label{subsec: 2a_6_output_buffer}
An additional component required by our emulation environment is the \textit{output buffer}.
In Compass\cite{Compass}, cores designated with output neurons send packets to an output buffer. This buffer is used to retrieve all output spike packets and send them to the user in a single \textit{tick} instance, rather than the user receiving spike packets at irregular intervals.
This extra buffer component adds an extra \textit{tick} instance between the outputs of a TrueNorth core and the user.
To ensure the output timings match up with the Compass results, we constructed a simple output buffer component that is attached to the emulated network.
Output spikes are retrieved by this component, accumulated during a \textit{tick} instance, and then sent to the user at the start of the next \textit{tick} instance.

\section{Verification and Scalability}\label{sec: 04_VERIFICATION}
In this section, we first present hardware setup and FPGA resource utilization for our TrueNorth prototype. Next, we present our approach of functional verification between our FPGA prototype and IBM's TrueNorth simulator, Compass \cite{Compass}, by implementing a 9-bit signed vector-matrix multiplication (VMM) algorithm. 
We then perform functional verification by comparing results on the MNIST dataset with published TrueNorth based implementation results~\cite{NIPS_2015_Training}.

\subsection{Streaming Framework}\label{subsec: streaming}
We use the Xilinx Zynq Ultrascale+ MPSoC (XCZU9EG2FFVB1156) as the implementation platform, which consists of a Programmable System (PS) of Quad-core ARM Cortex-A53 CPU integrated with a Programmable Logic (PL). The neuron parameters for the core SRAM, neuron instructions of each core, and the spikes are first loaded into the emulator. These three files are generated offline for deploying the network on the FPGA after all constraining and training has been completed using the \textit{constrain-then-train} methodology described by Esser et al. \cite{PNAS_2016_Applications}. In order to move data in and out of our platform, we utilize the DMA functionality of the Zynq SoC. The emulation utilizes both the PS and PL resources of the MPSoC platform. Overall architecture of the emulation environment is composed of two ARM CPU cores on the PS side acting as the host threads, and on the PL side, a DMA engine, buffer, and the TrueNorth module. The host thread of the first ARM CPU core reads packets from a binary file, which can be shared using an SD card. The packets are then written to shared memory, which the PL can access using the DMA engine. The packets are fed from memory into the TrueNorth module, where they are buffered to be read at each tick. While running, the output packets are also buffered, and at the end of each tick they are written to a separate part of the shared memory. The second ARM core reads from the shared memory and writes the output packets back to the SD card.

\subsection{Hardware Implementation Results}

Table \ref{tab: Resource_Requirements_1and5} shows the resource utilization and timing analysis on the Zynq Ultrascale+ for two network sizes. We use the single  and five core implementations for functional verification  against the VMM and MNIST reference designs respectively.  
The single core occupies 0.72\% of the logic resources, 0.21\% of the logic-memory resources, and 0.60\% of the BRAM resources. 
When we scale the network to the five cores, resource utilization increases linearly, occupying 3.88\% of logic resources, 1.06\% of logic-memory resources, and 3.02\% of the BRAM resources. 
Core computations involve 9-bit signed weight addition in the neuron block, along with 9-bit signed increment and decrement operations in the router block. 
Each core operates at global tick rate of 1KHz~\cite{TCAD_2015_Foundation}.

We show the resource utilization trend with respect to increase in the number of cores in Table~\ref{tab: FPGAUsage}. We sweep the resource usage space by beginning with our single core implementation. We then expand out in the \textit{x} and \textit{y} directions of our 2D network grid maintaining a square network. We observe that the network scales in a seemingly linear fashion, with the primary resource demands being on the LUT and BRAM components. We are able to create a 110 core (10x11) network, bounded by LUT utilization, as a rectangular grid on the  Zynq Ultrascale+ XCZU9EG. 
\begin{table}[t]
    \caption{Hardware resource usage with respect to the number of emulated TrueNorth cores. LUTs determine the scalability, reaching nearly 98\% utilization at 110 cores.}
    \centering
    \begin{tabular}{|c|c|c|c|c|}
    \hline
    NETWORK & LUT & LUTRAM & FF  & BRAM \\ 
    SIZE & (\%) & (\%) & (\%) & (\%)\\ \hline
    1    & 8.40  & 0.31     & 3.15  & 0.60   \\
    4    & 9.99  & 0.73     & 3.68  & 1.81   \\
    9    & 14.03 & 1.79     & 5.05  & 4.82   \\
    16   & 19.75  & 3.27     & 7.01  & 9.05   \\
    25   & 27.15 & 5.17     & 9.55  & 14.47  \\
    36   & 36.24 & 7.49     & 12.68  & 21.11  \\
    49   & 47.01 & 10.23    & 16.40 & 28.95  \\
    64   & 59.47 & 13.40    & 20.70 & 37.99  \\
    81   & 73.62 & 16.99    & 25.59 & 48.25  \\
    100  & 89.45 & 21.00    & 31.07 & 59.70  \\
    110 & 97.78 & 23.11 &      33.95 & 65.73 \\
    
    \hline
    \end{tabular}
    \label{tab: FPGAUsage}
    \vspace{-4mm}
\end{table}

\begin{table*}

\caption{Table depicts the tick by tick interaction between incoming spikes, connection weights, neuron potentials and output spikes depending upon axon type. We see that the symmetry of the reset thresholds affects the state of the neuron potential after successive ticks, with the asymmetric potential remaining negative until a positive value drives it back toward zero. In applications like VMM, the positive (+) and negative (-) representations rely on identical behavior of positive and negative potential resets to allow simultaneous positive and negative values to be calculated and represented by the neurons. Due to the configurability of our emulation environment, the feedback system used to correct the asymmetry is easily discarded.}
\centering

\begin{tabular}{cccc|c|c|c|c|c|c|c|c|}
\cline{5-12}
\multicolumn{1}{l}{}       & \multicolumn{1}{l}{}      & \multicolumn{1}{l}{}       & \multicolumn{1}{l|}{}              & \multicolumn{4}{c|}{Asymmetric}                                                                                                             & \multicolumn{4}{c|}{Symmetric}                                                                                                              \\ \hline
\multicolumn{1}{|l|}{Tick} & \multicolumn{1}{l|}{Axon} & \multicolumn{1}{l|}{Spike} & \multicolumn{1}{l|}{Weight(+, -)} & \multicolumn{1}{l|}{Potential(+)} & \multicolumn{1}{l|}{Potential(-)} & \multicolumn{1}{l|}{Output(+)} & \multicolumn{1}{l|}{Output(-)} & \multicolumn{1}{l|}{Potential(+)} & \multicolumn{1}{l|}{Potential(-)} & \multicolumn{1}{l|}{Output(+)} & \multicolumn{1}{l|}{Output(-)} \\ \hline
\multicolumn{1}{|c|}{1}    & \multicolumn{1}{c|}{0}    & \multicolumn{1}{c|}{1}     & 1, -1                              & 1                                  & -1                                 & 1                               & 0                               & 1                                  & 1                                  & 1                               & 0                               \\
\multicolumn{1}{|c|}{2}    & \multicolumn{1}{c|}{X}    & \multicolumn{1}{c|}{0}     & X                                  & 0                                  & -1                                 & 0                               & 0                               & 0                                  & 0                                  & 0                               & 0                               \\
\multicolumn{1}{|c|}{3}    & \multicolumn{1}{c|}{1}    & \multicolumn{1}{c|}{1}     & -1, 1                              & -1                                 & 0                                  & 0                               & 0                               & -1                                 & 1                                  & 0                               & 1       \\ \hline                       
\end{tabular}
\label{tab: spike_interaction}
\vspace{-4mm}
\end{table*}

\subsection{Vector Matrix Multiplication Verification}\label{subsec: VMM}

As shown by Fair et al.~\cite{fair19}, the mapping of vector-matrix multiplication (VMM) onto TrueNorth using Compass\cite{Compass}, IBM's TrueNorth simulator, spreads computation across network, core and neuron components, and  
 runs for hundreds of ticks. Therefore, it is an ideal application for testing our emulation environment. Furthermore, VMM has also proven to be a core building block in implementing multiple sophisticated algorithms on TrueNorth such as the locally-competitive algorithm \cite{fair19}, Word2Vec word similarity calculation \cite{Word2Vec}, and the neural engineering framework \cite{NEF}. 
We use the 9-bit signed VMM to verify the behavioral functionality of our emulation prototype against its implementation in Compass.
We created 100 random matrices ranging from 2x3 to 8x8 and a random vector of 9-bit signed integers of the appropriate size for each matrix. 
Each vector-matrix pair was mapped to Compass \cite{Compass} using the method proposed by Fair~\cite{fair19}. We then mapped the same matrix-vector pairs to our FPGA emulation and found a one-to-one match between the two. 

\begin{table}[]
\centering
\caption{Post implementation resource utilization when running a 9-bit signed VMM implementation on a single core. The first row represents the core before architectural modification, while the second row represents the core after the neuron behavior modification and removal of feedback system.
}
\label{tab:VMMmapping}
\begin{tabular}{|c|ccccc|}
\hline
Design         & LUT  & LUT-RAM & FF   & BRAM & Delay (ns) \\ \hline
Reference      & 1700 & 192     & 1210 & 4  & 10.266     \\
Proposed       & 1165 & 48     & 1056 & 2  & 8.026      \\ \hline
Reduction (\%) & 31.5 & 75.0    & 12.7  & 50.0 & 21.8       \\ \hline
\end{tabular}
\vspace{-2mm}
\end{table}

\subsection{Modifying Neuron Behaviour for Efficient VMM Mapping}
Mapping signed VMM requires representation of positive and negative values. As the rate encoded input spikes lack sign, to make signed VMM mappable to the TrueNorth, Fair et al. \cite{fair19} duplicates the axons in a core, dividing them into positive and negative groups, where positive and negative input spikes are routed to their respective groups. Similarly, the neurons are duplicated and divided, allowing them to represent positive and negative outputs from respective connected axons. Neuron block operations proceed as previously described, facilitating simultaneous positive and negative value operation without previous knowledge of sign.

While the positive threshold is evaluated using $\geq$, the negative threshold uses the $<$ operator as illustrated in Figure~\ref{fig: neuron_block}. 
Uncorrected, this asymmetry allows the potential of a neuron to remain negative when it should otherwise be reset to zero, thus producing an incorrect number of output spikes. This is depicted in Table~\ref{tab: spike_interaction}. Despite the neuron and axon duplication that allows signed VMM, correct output can only be achieved by implementing a feedback system that reroutes spikes back to the core and drives the negative neuron potential back to zero~\cite{fair19}. This feedback system requires an additional doubling of the number of neurons.   

The scalability of VMM mapping to TrueNorth is limited by this asymmetry of neuron potential reset thresholds. The duplication of the number of axons and neurons severely limits scalability, quickly exhausts resources on TrueNorth for larger VMM problems and requires a cluster of TrueNorth chips to map convolution, locally competitive algorithm, least squares minimization, or support vector machine training~\cite{fair19} types of applications.
Resorting to a resource replication type of workaround to accommodate signed multiplication is inevitable when restricted by the fixed architecture. We identify this problem as a key case study for demonstrating the utility of our FPGA based emulation environment, where an application engineer has the ability to change the hardware behavior and eliminate the need for resource duplication completely. In our emulation environment, the imbalance of equality operators is quickly resolved by modifying the negative threshold behavior such that it uses a "$\leq$" comparison rather than a "$<$" comparison. Despite the simplicity of this change in hardware, it is infeasible within Compass due to the fixed-nature of the TrueNorth architecture. 
The proposed symmetric threshold based hardware modification eliminates the need for the feedback system, and it enables resource reduction,which we discuss next.
 
To demonstrate the results, we map an $8\times8$ matrix, the largest which fits a single $256\times256$ core with feedback. Each column requires 8 neurons for the positive representation, 8 neurons for the negative representation, and 16 neurons for feedback. This is 32 neurons per column, multiplied by 8 columns, yielding 256 neurons. 16 feedback neurons multiplied by 8 columns, yields 128 feedback neurons, which necessitate 128 axons by which to connect.
The maximum vector of $1\times8$ requires 1 axon per column for positive and 1 axon per column for negative representation. Duplicating this number ensures correspondence between signed inputs and signed matrix values resulting with 4 axons per column and a total of 32 axons. The 128 feedback axons, required by the $8\times8$ matrix, are added to the 32 input axons, creating a 160 axon, 256 neuron core. Eliminating the feedback system, leaves behind a 32 axon, 128 neuron core to solve the same VMM problem and reduces the number of neurons by 50\%. 

In order to validate our resource reduction analysis, we implement the signed VMM mapping method of Fair et al.~\cite{fair19} for the $1\times8$ vector and $8\times8$ matrix on the reference architecture that is emulated using the Zynq Ultrascale+ MPSoC. We then implement the same VMM problem on the proposed architecture that supports symmetric threshold and eliminates the feedback loop. We show the resource usage of the functionally equivalent VMM mappings on the reference and proposed architectures in Table~\ref{tab:VMMmapping}.
Elimination of the feedback loop removes half the dimensions of a standard core, in turn reduces the necessary bit allotment for the \textit{scheduler} and \textit{core sram}. 
As shown in Table~\ref{tab: FPGAUsage}, the standard $256x256$ core requires 5.5 BRAMs, which the \textit{core sram} occupies. For the $8\times8$ VMM problem, based on the reference design with the feedback loop, we note the smaller dimensions of $160\times256$, requires 4 BRAMs. For the proposed symmetric threshold based architecture, the $32\times128$ core requires 2 BRAMs, which confirms the expected 50\% reduction. 
Instead of an 8-bit counter for the \textit{scheduler} that counts up to 256, the proposed design requires a 5-bit counter to index each of the 32 axons. This enables more efficient mapping of the scheduler to the LUT-RAMs, reducing total utilization by 75\%. Additionally, removing the feedback system not only permits a matrix capacity double the previous to occupy a single core, but also reduces the critical path delay by 21.8\%, further bolstering throughput and scalability.

\subsection{MNIST Verification}\label{subsec: Five_Core_Network_Verification}
\begin{figure}[t]
    \centering
    \includegraphics[width=0.75\linewidth]{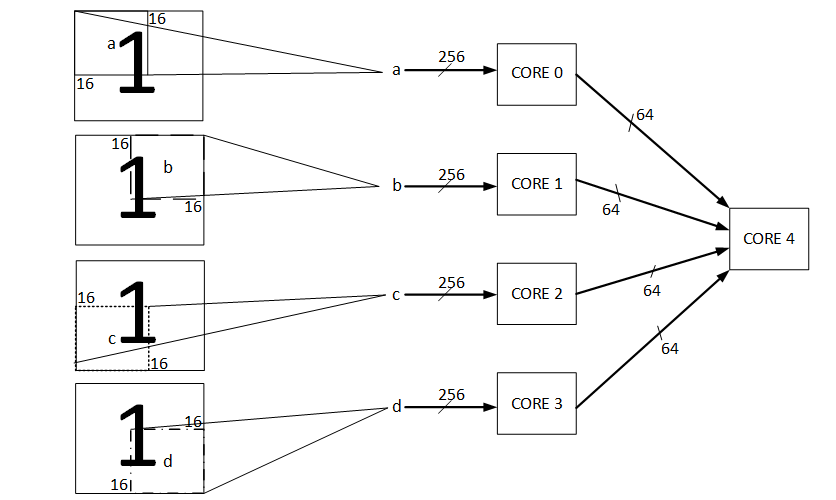}
    
    \caption{5 core network implementation with four 16x16 windows with a stride of 12 represented by blocks a-d. These generate a 256 input fan-in for the input layer of cores in our network. Input layer only uses 64 of the 256 possible neurons and outputs those to the classification core in the next layer.
    }
    \label{fig: five_core_network}
    \vspace{-4mm}
\end{figure}
In this experiment we implement a five core network that replicates the design introduced by Esser et al. \cite{NIPS_2015_Training}, illustrated in Figure \ref{fig: five_core_network}, while using the training methodology proposed by Yepes et al. \cite{IJCAI_2017_Training}. 
Due to the MNIST data set images having a size of 28x28 pixels, to ensure that our input layer of cores are fully connected, we split the images into four sections of 16x16 windows with each window separated using a stride of 12 pixels. 
Each core of our input layer only uses 64 out of the 256 total neurons available to them, as each input core represents one-quarter of the image. 
The image splitting method ensures each quarter is weighted evenly within the classification core.
The classification core then uses only 250 out of the total 256 neurons to evenly distribute 25 neurons per class for each of the ten classes within MNIST data set.

\begin{figure}[t]
    \centering
    \includegraphics[width=0.85\linewidth]{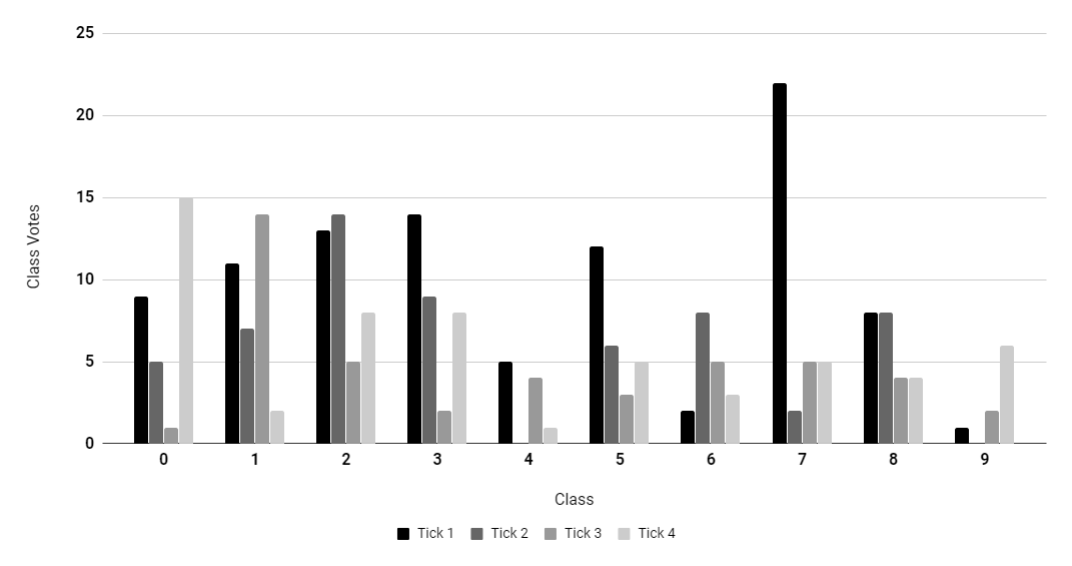}
    \caption{For the MNIST data set we modify our \textit{Output Core} to output all class votes as they are accumulated. For the first six \textit{ticks} of the data set, we generate the resulting votes in the above histogram. For instances where a tie occurs, the \textit{Output Core} is set to select the first instance.}
    \label{fig: MNIST_Hist}
    \vspace{-5mm}
\end{figure}

When running our implementation, the output core generates all votes for each class in the MNIST as they are accumulated. This allows us to produce a histogram similar to Figure~\ref{fig: MNIST_Hist}, which shows the number of votes for each digit across multiple ticks. By comparing against Compass we verified that these histograms were matching and the correct digit was being selected. Our five core implementation achieves an accuracy of 96.28\% on the MNIST data set, which is comparable to the accuracy achieved by Yepes et. al \cite{IJCAI_2017_Training}. Our emulation environment takes 10 seconds to fully infer the 10,000 testing images of the MNIST. An in-house serial implementation of the same emulation environment takes around 2 hours on an Intel Xeon processor (3GHz, 32GB RAM) to fully infer the dataset. We find that even with the symmetric threshold, our accuracy is unchanged.
\vspace{-1mm}

\section{Conclusion}\label{sec: 05_CONCLUSION}
In this paper we present our approach to implementing an FPGA-based neuromorphic architecture emulation platform.  
We use IBM’s TrueNorth as a reference and discuss our hardware design decisions for each architectural component to make it feasible to implement on the FPGA.  
We conduct hardware resource usage analysis, validate the functionality of our emulation environment and demonstrate its utility through case studies based on comparisons with respect to the published results.  
To the best of our knowledge this is the first academic work on FPGA based emulation environment for simulating the clusters of leaky-integrate-and-fire neuron models integrated with the principle router, scheduler, and memory management components. Unlike other approaches (e.g.,~\cite{neuron97,genn16, carlsim18, simulation13}) that are presented towards achieving large scale spiking neural network simulations, the proposed open-source, parameterized and modular emulation environment serves as a basis to conduct hardware architecture research for neuromorphic computing and investigate the trade space between mapping strategies, hardware performance and accuracy for the target applications.

Our FPGA-based emulation environment 
replaces the "globally synchronous-locally asynchronous" design with a fully synchronous design as we focus on designing a functionally correct synaptic cores and basic leaky-integrate-and-fire neuron model. This allowed us to rapidly manipulate core components without needing to continually reconfigure our FPGA place-and-route tool chains to meet asynchronous timing requirements, and investigate applications which have difficulty being mapped due to the architectural constraints, as we demonstrated with the case study on VMM requiring neuron copies to correctly function.

We believe there is room for reducing the resource usage for a better salable emulation platform. 
We plan to optimize the BRAM usage by replacing the method of reading all 386 bits in a single clock cycle with a design that reads from the core sram in 72 bit bursts over multiple clock cycles aligned with the 512x72 BRAM configuration.
As future work we plan to build on our resource efficient way of mapping the VMM and implement applications such as sparse matrix approximation and convolution. 
The ability to process convolution in turn will allow us to target a much broader class of image recognition tasks, such as Synthetic Aperture Radar (SAR) classification; dealing with more complex images compared to MNIST. 
Additionally, we will investigate the model accuracy challenges of a neuromorphic system while maintaining its energy efficient execution flow by studying correlation between training methods, accuracy, and architecture configuration parameters.

\section{Acknowledgements}\label{sec: 06_ACKNOWLEDGEMENTS}
Research reported in this publication was supported in part by Raytheon Missile Systems under the contract 2017-UNI-0008. 
The content is solely the responsibility of the authors and does not necessarily represent the official views of Raytheon Missile Systems.

\bibliographystyle{abbrv}
\bibliography{references}
\end{document}